\documentclass{article}
\usepackage[T1]{fontenc} 
\usepackage[utf8]{inputenc} 
\usepackage{ismir,amsmath,cite,url}
\usepackage{graphicx}
\usepackage{color}
\usepackage[dvipsnames]{xcolor}
\usepackage{booktabs}
\usepackage{tabularx} 
\usepackage{multirow}
\usepackage{makecell}
\usepackage{amssymb,amsmath,mathtools}
\usepackage{amsthm}
\usepackage[ruled,vlined,linesnumbered,resetcount]{algorithm2e}
\usepackage{url}
\usepackage{siunitx}

\usepackage{harmony}

\usepackage[normalem]{ulem}
\useunder{\uline}{\ul}{}

\usepackage{lineno}

\title{On the Characterization of \\
Expressive Performance in Classical Music: \\
First Results of the \textit{Con Espressione} Game}

\multauthor
{Carlos Cancino-Chac\'on$^{1,4}$ \hspace{1cm} Silvan Peter$^2$ \hspace{1cm} Shreyan Chowdhury$^2$} { \bfseries{Anna Aljanaki$^3$ \hspace{1cm} Gerhard Widmer$^{1,2}$}\\
$^1$ Austrian Research Institute for Artificial Intelligence, Vienna, Austria\\
$^2$ Institute of Computational Perception, Johannes Kepler University Linz, Austria\\
$^3$  Institute of Computer Science, University of Tartu, Estonia\\
$^4$ RITMO Centre for Interdisciplinary Studies in Rhythm, Time and Motion, University of Oslo, Norway\\
{\tt\small carlos.cancino@ofai.at, silvan.peter@jku.at}
}
\def\authorname{C.~Cancino-Chac\'on, S.~Peter, S.~Chowdhury, A.~Aljanaki and G.~Widmer}

\begin{document}

\maketitle
\begin{abstract}
A piece of music can be expressively performed, or interpreted, in a
variety of ways. With the help of an online questionnaire, the
\textit{Con Espressione} Game, 
we collected some 1,500 descriptions of expressive character
relating to 45 performances of 9 excerpts from classical piano pieces,
played by different famous pianists. More specifically, listeners
were asked to describe, using freely chosen words (preferably:
adjectives), how they perceive the expressive character of the different
performances. 
In this paper, we offer a first account of this new data resource for
expressive performance research, and provide an exploratory
analysis, addressing three main questions:
(1) how similarly do different listeners describe a performance of
a piece?
(2) what are the main dimensions (or axes) for expressive
character emerging from this?; and
(3) how do measurable parameters of a performance (e.g., tempo,
dynamics) and mid- and high-level features that can be predicted
by machine learning models (e.g., articulation,
arousal) relate to these expressive dimensions?
The dataset that we publish along with this paper was enriched by adding hand-corrected
score-to-performance alignments, as well as descriptive audio features
such as tempo and dynamics curves.

\end{abstract}
%


\section{Introduction}\label{sec:introduction}

In the Western classical music tradition, music exists at an interplay of creative intentions of  composer, performers and listeners. 
Composers encode their ideas using written notation (i.e., musical scores), and performers bring these ideas to life, guided by the expression markings, performance traditions, and their  own creative imagination. 
Each performance can sound very different from the next.

Much of the research on musical expression has focused on what pieces express through attributes of their musical structure~\cite{Davies:2001uw,Rosen:2010,JuslinSloboda_PoM_Chapter:2013}.
There has been much focus on the expression of \textit{emotion} in particular~(e.g., \cite{Juslin:wg,juslin97:_emotion_commun_music_perfor,juslin2010expression,Juslin2003,doi:10.1177/10298649020050S104}); however, a comprehensive description of the expressive character of a performance includes additional, not specifically emotion-related aspects. 
The aim of this research is to find the dimensions of musical expression that can be attributed to a performance, as perceived and described in natural language by listeners. 

Within the classical music tradition, there is already a practice of assigning verbal descriptors to aspects of musical expression. 
For example, instructions related to expressive character are sometimes marked on the score by the composer with a fixed set of (mostly) Italian terms (e.g., 
\textit{Allegro}, \textit{dolce}). 
Many of those terms describe emotions, but they describe a wide range of other aspects as well, including movement analogies and metaphors (e.g., free, flowing) \cite{sulem2019perception}. 

Using an online questionnaire, the \emph{Con Espressione} Game (CEG), which is part of the research project of the same name~\cite{widmer:manifesto:2017}, we collected verbal descriptors of expressive performances.
In this paper we present first results on three main questions: 
(1) can we observe any consistency in the way listeners perceive and describe \emph{expressive character}?; 
(2) can we identify main descriptive dimensions along which these characterizations can be organized?; and
(3) how are these dimensions related to measurable qualities (or parameters) from audio recordings, or to performance information extracted from these?


The rest of this paper is structured as follows:
Section \ref{sec:related_work} points to some related work.
Section \ref{sec:con_espressione_game} describes the CEG and the collected data.
We continue the paper with individual sections addressing the three main questions we want to study:
Section \ref{sec:q1_similarity} presents results on how similar/consistent the characterizations of the participants are,
Section \ref{sec:q2_dimensions} focuses on analyzing the main descriptive dimensions.
Section \ref{sec:q3_performance} presents results relating performance and audio features to the expressive dimensions.
Section \ref{sec:discussion} discusses the results of these experiments.
Finally, Section \ref{sec:conclusions} concludes the paper.

\section{Related Work}\label{sec:related_work}

\begin{table*}[t]
\begin{small}
 \begin{tabular}{llrl}
  \hline
  Composer & Piece & \# & Pianists \\
  \hline
  Bach & Prelude No.1 in C, BWV 846 (WTC I) & 7 &
  Gieseking, Gould, Grimaud, Kempff, Richter, Stadtfeld, MIDI \\
  Mozart & Piano Sonata K.545 C major, 2nd mvt. & 5 &
  Gould, Gulda, Pires, Uchida, MIDI \\
  Beethoven & Piano Sonata Op.27 No.2 C\# minor, 1st mvt. & 6 &
  Casadesus, Lazi\'c, Lim, Gulda, Schiff, Schirmer \\
  Schumann & Arabeske Op.18 C major (excerpt 1) & 4 &
  Rubinstein, Schiff, Vorraber, Horowitz \\
  Schumann & Arabeske Op.18 C major (excerpt 2) & 4 &
  Rubinstein, Schiff, Vorraber, Horowitz \\
  Schumann & Kreisleriana Op.16; 3. Sehr aufgeregt (ex.~1) & 5 &
  Argerich, Brendel, Horowitz, Vogt, Vorraber \\
  Schumann & Kreisleriana Op.16; 3. Sehr aufgeregt (ex.~2) & 5 &
  Argerich, Brendel, Horowitz, Vogt, Vorraber \\
  Liszt & Bagatelle sans tonalit\'e, S.216a & 4 &
  Bavouzet, Brendel, Katsaris, Gardon \\
  Brahms & 4 Klavierst\"ucke Op.119, 2. Intermezzo E minor & 5 &
  Angelich, Ax, Serkin, Kempff, Vogt \\
 \hline
 \end{tabular}
\end{small}
 \label{tab:pieces}
 \caption{Performances used in the \emph{Con Espressione} Game.}
\end{table*}

In this section we provide a few pointers to related work on research on music emotion and expressive performance, and on characterizing musical expression.
A full overview is beyond the scope of this paper.

From a musicological and philosophical viewpoint, the study of musical expression has focused on music as in composition~\cite{Davies:2001uw,Rosen:2010}, although there is a trend in recent times to include and recognize both the role of the performer and the role of the listener.
Some recent papers have focused on developing mid-level features that capture qualities of musical recordings that relate to perceptual aspects of the music \cite{friberg2014using, aljanaki2018data} .
For an overview of literature on emotion in music, we refer the reader to \cite{JuslinSloboda_PoM_Chapter:2013} and \cite{yang2012machine}.

Computational models of expressive performance are a means to
study principles of performance in quantitative terms.
In Western classical music, much of this work has focused on establishing relationships between structural aspects of a musical score and quantifiable aspects of a performance such as expressive timing and dynamics.
At the intersection of research on music emotion and computational models of performance is the study of the relation between performance parameters (timing, dynamics) and emotion~\cite{Juslin:wg,juslin97:_emotion_commun_music_perfor,juslin2010expression,Juslin2003,doi:10.1177/10298649020050S104}.
For an overview of computational modeling of music performance, we refer the reader to~\cite{CancinoChacon:2018po}.

Particularly relevant to the current study are adjective lists for describing musical expressiveness, including the seminal work by Hevner~\cite{Hevner:1936} and its more recent updates ~\cite{Farnsworth:1954,Schubert:2003}.
Hevner identified several clusters of adjectives that describe musical expression.
Many of these clusters relate to emotions, and include aspects such as \emph{`happy'} and \emph{`sad'}.
Schaerlaeken et al.
\cite{SchaerlaekenEtAl:2019} conducted a large study to investigate the use of metaphors for describing Western classical music.
They propose the Geneva Musical Metaphors Scale (GEMMES), which comprises 5 metaphorical scales including aspects like \emph{flow, movement, force, interior} and \emph{wandering}.
The focus of their study was on characterization of different pieces of music, rather than on description of the character of different expressive performances.
A related study
\cite{sulem2019perception} presents a perception-based clustering of expressive musical terms (i.e., performance directives such as \textit{catabile} and \textit{leggiero}), relating these terms to locations in Russell's arousal--valence space~\cite{Russell:1980wy}.
Murari et al.~\cite{doi:10.1080/09298215.2015.1101475} study of listeners' characterization of music using non-verbal sensory scales.



\begin{figure}[t]
\centerline{\includegraphics[width=0.85\columnwidth]{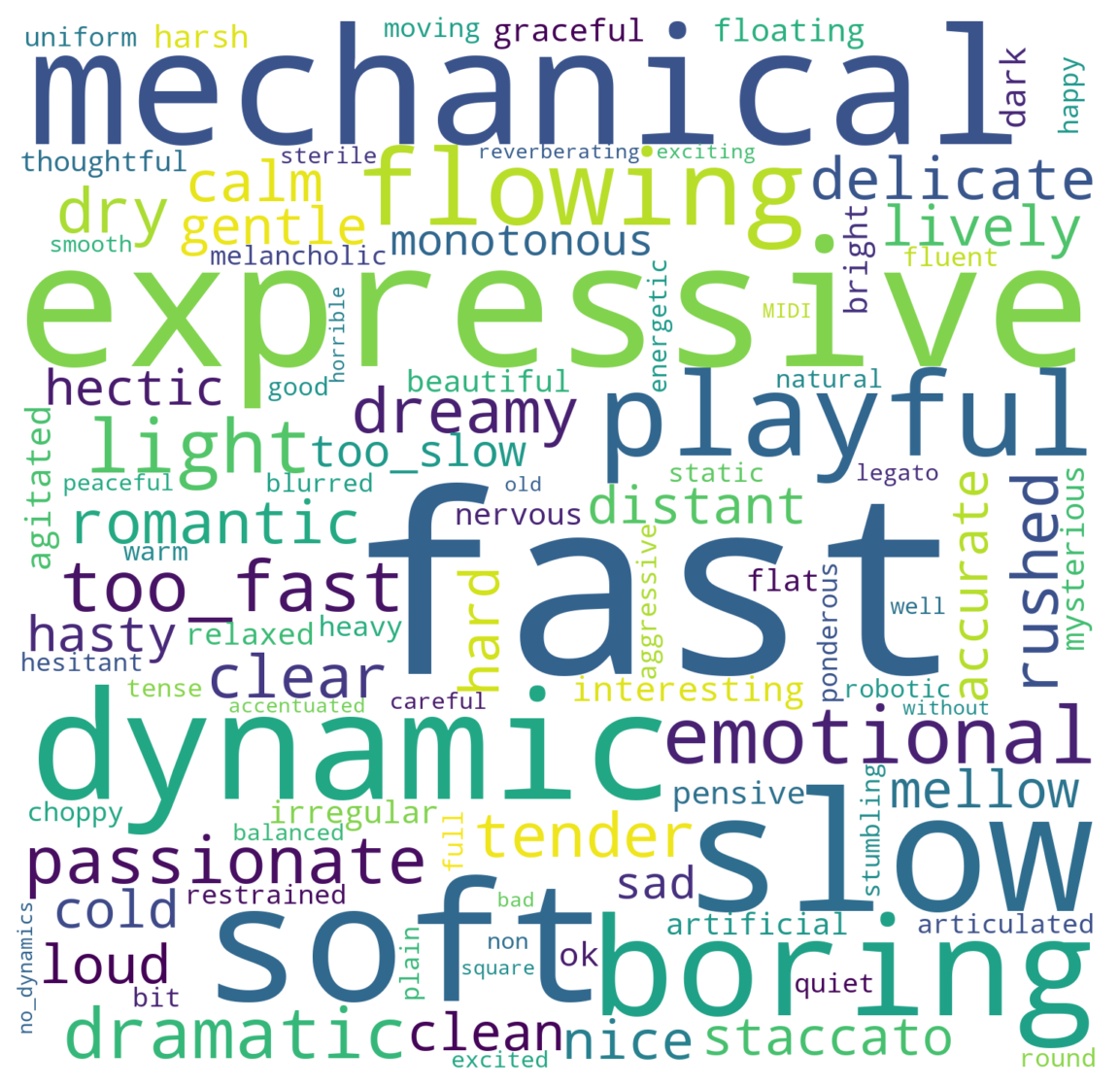}}
\caption{A word cloud of the terms in the dataset.}
\label{fig:wordcloud}
\end{figure}

\section{The \emph{Con Espressione} Game}\label{sec:con_espressione_game}

We collected our data through a Web-based questionnaire (``CEG''), 
where participants listened to several performances of each of 9 classical piano pieces.
They were then asked to describe the performance in free text (preferably adjectives, as many as they liked), concentrating on the performative aspects and not on the piece itself. The target phenomenon thus is what we would call the \emph{expressive character} of performances. 
Users could also select their favorite performance of each piece.
Upfront, a few general questions were asked regarding their
level of musical education, how often they listen to classical music, and whether they
play the piano.
The order of pieces and performances was randomized. 
Users could stop at any time.


\subsection{Pieces and Responses}


 
The CEG contains 45 performances of 9 excerpts (see Table \ref{tab:pieces}).\footnote{\url{con-espressione.cp.jku.at}}
The length of the excerpts was between 27 seconds and 188 seconds. 
The CEG was launched on the 3rd of April 2018. 
The questionnaire was filled out by 194 participants, out of which 88\% had some kind of music education -- on average 11.7 years;
179 participants answered in English, 12 in German and three  in each of Russian, Spanish and Italian.
On average, participants listened to the performances of 4.5 out of 9 pieces, 27 participants listened to all the 45 excerpts.


\subsection{Meta-data and Annotations}
We enriched the dataset by adding score-to-performance alignments for all performances in the dataset,
as well as descriptive performance features.
The alignments were produced by manually annotating the position of the beats in the audio files using Sonic Visualiser~\cite{SonicVisualiser}.
The scores were encoded manually and exported in MusicXML format following publicly available editions.\footnote{from \url{https://imslp.org}.}
From these alignments we compute performance features such as tempo and loudness curves (see Section \ref{sec:audio_performance_features}).
The dataset is available online.\footnote{The dataset and code necessary to reproduce the experiments reported in this paper can be found in \url{https://cpjku.github.io/con_espressione_game_ismir2020/ }.}

\section{How similarly do listeners describe a performance of a piece?}\label{sec:q1_similarity}

Probably the first question that arises 
concerns the similarity of the descriptions in the dataset, i.e., whether there are commonalities in the way listeners describe and like performances.
In this section we present a series of analyses that provide different perspectives on the data. 




\subsection{Frequency and Distribution of Terms}\label{sec:dataset_statistics}

Users provided 1,515 individual descriptions for a total of 3,166 terms, of which 1,415  (approx.~45\%) are unique.
Figure \ref{fig:wordcloud} shows a word cloud of the terms appearing in the dataset.
Taken all answers together, each performance was described using at least 47 and at most 114 terms.
The average number of terms per piece varies between 60.3 for Liszt's \textit{Bagatelle sans tonalit\'e} and 98.4 for Bach's \textit{Prelude in C}.
Each performance is characterized by at least 44 and at most 91 unique terms.
The average number of unique terms per piece varies between 51.0 for Liszt's Bagatelle and 78.4 for Bach's Prelude.

The terms \emph{`dynamic'}, \emph{`expressive', `fast', `like', `loud', `mechanical', `slow'}, and \emph{`soft'} appear in at least one performance of every piece.
420 terms appear more than once and, coincidentally, also for different performances.
Only 40 terms appear in more than ten performances.
The most frequently used terms \emph{across performances} are \emph{`dynamic', `expressive'}, and \emph{`soft'}, which appear in at least 22 performances.
In increasing order of occurrence, the terms \emph{`playful', `boring', `dynamic', `mechanical', `slow', `soft', `expressive'}, and \emph{`fast'} are used at least 40 times, with this last term being the most used with 64 occurrences.




\subsection{Complexity of the Descriptions}\label{sec:readability}

An interesting question is whether there is a relation between listeners' musical backgrounds and the complexity of their answers.
In particular, we are interested in determining whether listeners with more musical experience with Western classical music describe performances of this kind of music in a more complex language.
As a measure of the complexity of the descriptions we use the Dale-Chall readability score\cite{DaleChall:1948}, a
measure that takes into account the number and commonality of words (
defined as words that would be familiar to American students) in a weighted sum. 
Intuitively, we can understand this measure as a way of combining the length (i.e., number of terms) of the answer and the use of specialized vocabulary (such as musical terms). 
A larger score means a more complex answer. 
To test whether listeners with more musical training describe performances in a more complex language, we conduct a linear regression (years of musical training vs.~answer complexity).
This analysis reveals a positive correlation that we assessed for statistical significance using a one-tailed Wald test (Pearson's $r=0.27$, $W = 2.32$, $p=0.01$, $R^2 = 0.07$), which suggests a small effect.
We also test whether listeners that often \textit{listen} to classical music describe performances in more complex terms, and find a small, non-significant correlation.
These results present weak evidence supporting the idea that listeners with more musical experience describe the expressive character in more complex ways.

\subsection{Listeners' Preferences}\label{sec:listeners_preference}

To determine how similar the preferences of the listeners are, i.e., if they like (or dislike) similar performances we compute a $\chi^2$-test for each piece
to determine if there are clear preferred performances, or if all performances of the same piece are equally liked.
There are only three pieces that reject the null hypothesis (the frequency of preferred performances is flat) at a level $\alpha = 0.01$, namely the performance of Bach's piece by Richter; and the performances of the two excerpts of Schumann's Arabeske  (in this case, the preferred performances were both by Vorraber).
Two additional pieces reject the null hypothesis at level $\alpha = 0.05$: Schumann's Kreisleriana performed by Brendel and Mozart's  piece by Pires.
Both Gould's performances and the deadpan MIDI performances (for both Bach and Mozart) were the least preferred for this piece.
It is important to emphasize that this analysis only indicates the presence of preferred performances; it does not identify what the preferred performance is.

\subsection{Semantic Similarity of the Descriptions}\label{sec:semantic_similarity}

We use Li et al.'s~\cite{LiEtAl:2006} method for estimating the semantic similarity of short sentences, to compute the pairwise similarity between descriptions of performances.
Similarity between individual terms is estimated as the semantic distance between words in WordNet~\cite{Wordnet}, and the overall sentence similarity is weighted with corpus statistics.
Intuitively, this method quantifies the overlap in terms (and very directly related synonyms of these terms) between the answers of the participants.
The average similarity of the descriptions of a performance of a piece by the same pianist (i.e., \emph{intra-performance}) is $16\%$,
the average similarity of descriptions of a performance and the descriptions of other performances of the same piece by different pianists (i.e., \emph{inter-performance}) is $15\%$;
and the average similarity of descriptions of a performance and performances of other pieces (i.e., \emph{inter-piece}, excluding performances of the same piece by other pianists) is $15\%$.
To test whether the differences between these groups are significant, we performed a one-way ANOVA ($F(132, 2) = 9.8$, $p<0.001$, $\eta^2 = 0.13$), which suggests a medium-small effect.
We use t-tests with Bonferroni correction to test the pairwise differences (3 tests, $\alpha=.02$).
These tests suggest that the average intra-performance similarity is larger than both the average inter-performance (one-tailed $t(88) = 3.4$, $p<0.001$, Cohen's $d=.76$) and inter-piece similarity (one-tailed $t(88) = 3.6$, $p< 0.001$, Cohen's $d=.71$).
The difference between the inter-performance and inter-piece similarities is not statistically significant (two-tailed $t(88) = 0.0$, $p= 0.99$, Cohen's $d=0.0$).
These results suggest that listeners describe a performance of a piece in more similar terms than they describe other performances of the same piece by different pianists.
A possible explanation for the fact that there is little variation in how listeners describe performances of different pieces could be that listeners have a limited vocabulary with which to distinguish the difference in expressive character.

\section{What are the main dimensions for expressive character?}\label{sec:q2_dimensions}



Most of the terms in Hevner's adjective checklist~\cite{Hevner:1936}, as well as the main five dimensions of expressive performance markings identified by Sulem et al.~\cite{sulem2019perception} and main factors of the GEMMES~\cite{SchaerlaekenEtAl:2019} are present in listeners' responses.
However, the characterizations go beyond the aforementioned clusters in at least three aspects:
many terms are non-emotional, technical, or disapproving.
Non-emotional terms are e.g., those that are hard to unambiguously place in the arousal--valence space (unlike Hevner's or Sulem et al.'s clusters) such as \emph{`clean'}, \emph{`metallic'} or \emph{`[t]his is Glenn Gould, obviously'}.
Technical terms include terms that describe playing techniques such as \emph{`legato', `staccato'}, or more generally \emph{`fast', `loud'}, and \emph{`mechanical'}.
Disapproving terms include descriptions with negative connotations such as \emph{`boring', `sterile'}, or \emph{`robotic'}.

Regarding automated analysis,
characterization of expressive performance is not a common case in natural language processing (NLP). 
Also, the meanings of many terms in the context of expressive performance are slightly different from their common usage.
Preliminary tests indicate that learned occurrence-based semantics on related or general topic corpora largely fail to represent more than superficial similarity for this dataset. 

In order to identify the main dimensions of terms, we compute a principal component analysis (PCA) on the occurrence matrix of the dataset.
The data is preprocessed in several steps:
answers in other languages are translated to English and terms are stemmed. 
A term is omitted if (1) it shows up less than three times in the dataset (its contribution to the global variance is minimal) or (2) it is used for all interpretations of the same piece (its piecewise entropy is zero; for instance, a participant wrote \emph{`i love mozart'} (sic) for all performances of the Mozart piece).

\begin{table*}[t]
\centering
\begin{small}
\begin{tabular}{@{}llllllll@{}}
\toprule
\multicolumn{4}{c}{{\color[HTML]{000000} \textbf{Dimension 1}}}                                                                    & \multicolumn{4}{c}{{\color[HTML]{000000} \textbf{Dimension 2}}}                                                                    \\
\multicolumn{2}{c}{{\color[HTML]{000000} positive loading}} & \multicolumn{2}{c}{{\color[HTML]{000000} negative loading}}  & \multicolumn{2}{c}{{\color[HTML]{000000} positive loading}} & \multicolumn{2}{c}{{\color[HTML]{000000} negative loading}}  \\ \midrule
{\color[HTML]{000000} hectic}     & {\color[HTML]{000000} 0.17} & {\color[HTML]{000000} sad}        & {\color[HTML]{000000} -0.20} & {\color[HTML]{000000} rushed}    & {\color[HTML]{000000} 0.22}  & {\color[HTML]{000000} hard}       & {\color[HTML]{000000} -0.19} \\
{\color[HTML]{000000} staccato}   & {\color[HTML]{000000} 0.15} & {\color[HTML]{000000} gentle}     & {\color[HTML]{000000} -0.18} & {\color[HTML]{000000} nervous}   & {\color[HTML]{000000} 0.20}  & {\color[HTML]{000000} stumbling}  & {\color[HTML]{000000} -0.18} \\
{\color[HTML]{000000} hasty}      & {\color[HTML]{000000} 0.15} & {\color[HTML]{000000} tender}     & {\color[HTML]{000000} -0.18} & {\color[HTML]{000000} too fast}  & {\color[HTML]{000000} 0.17}  & {\color[HTML]{000000} staccato}   & {\color[HTML]{000000} -0.17} \\
{\color[HTML]{000000} agitated}   & {\color[HTML]{000000} 0.14} & {\color[HTML]{000000} calm}       & {\color[HTML]{000000} -0.16} & {\color[HTML]{000000} bit}       & {\color[HTML]{000000} 0.16}  & {\color[HTML]{000000} ponderous}  & {\color[HTML]{000000} -0.14} \\
{\color[HTML]{000000} irregular}  & {\color[HTML]{000000} 0.14} & {\color[HTML]{000000} graceful}   & {\color[HTML]{000000} -0.16} & {\color[HTML]{000000} hasty}     & {\color[HTML]{000000} 0.15}  & {\color[HTML]{000000} monotonous} & {\color[HTML]{000000} -0.13} \\ \midrule
\multicolumn{4}{c}{{\color[HTML]{000000} \textbf{Dimension 3}}}                                                                    & \multicolumn{4}{c}{{\color[HTML]{000000} \textbf{Dimension 4}}}                                                                    \\
\multicolumn{2}{c}{{\color[HTML]{000000} positive loading}} & \multicolumn{2}{c}{{\color[HTML]{000000} negative loading}}  & \multicolumn{2}{c}{{\color[HTML]{000000} positive loading}} & \multicolumn{2}{c}{{\color[HTML]{000000} negative loading}}  \\ \midrule
{\color[HTML]{000000} monotonous} & {\color[HTML]{000000} 0.22} & {\color[HTML]{000000} heavy}      & {\color[HTML]{000000} -0.14} & {\color[HTML]{000000} ok}        & {\color[HTML]{000000} 0.24}  & {\color[HTML]{000000} cold}       & {\color[HTML]{000000} -0.15} \\
{\color[HTML]{000000} bad}        & {\color[HTML]{000000} 0.17} & {\color[HTML]{000000} graceful}   & {\color[HTML]{000000} -0.13} & {\color[HTML]{000000} happy}     & {\color[HTML]{000000} 0.21}  & {\color[HTML]{000000} warm}       & {\color[HTML]{000000} -0.14} \\
{\color[HTML]{000000} warm}       & {\color[HTML]{000000} 0.16} & {\color[HTML]{000000} smooth}     & {\color[HTML]{000000} -0.12} & {\color[HTML]{000000} joyful}    & {\color[HTML]{000000} 0.19}  & {\color[HTML]{000000} floating}   & {\color[HTML]{000000} -0.14} \\
{\color[HTML]{000000} peaceful}   & {\color[HTML]{000000} 0.16} & {\color[HTML]{000000} ponderous}  & {\color[HTML]{000000} -0.12} & {\color[HTML]{000000} free}      & {\color[HTML]{000000} 0.15}  & {\color[HTML]{000000} blurred}    & {\color[HTML]{000000} -0.14} \\
{\color[HTML]{000000} beautiful}  & {\color[HTML]{000000} 0.15} & {\color[HTML]{000000} soaring} & {\color[HTML]{000000} -0.10} & {\color[HTML]{000000} breathy}   & {\color[HTML]{000000} 0.14}  & {\color[HTML]{000000} mysterious} & {\color[HTML]{000000} -0.13} \\ \bottomrule
\end{tabular}

\end{small}
\caption{Terms with strongest loadings for each expressive character dimension.}
\label{tab:expressive_dimensions}
\end{table*}

\begin{figure*}[t]
\centering
\includegraphics[width=0.825\linewidth]{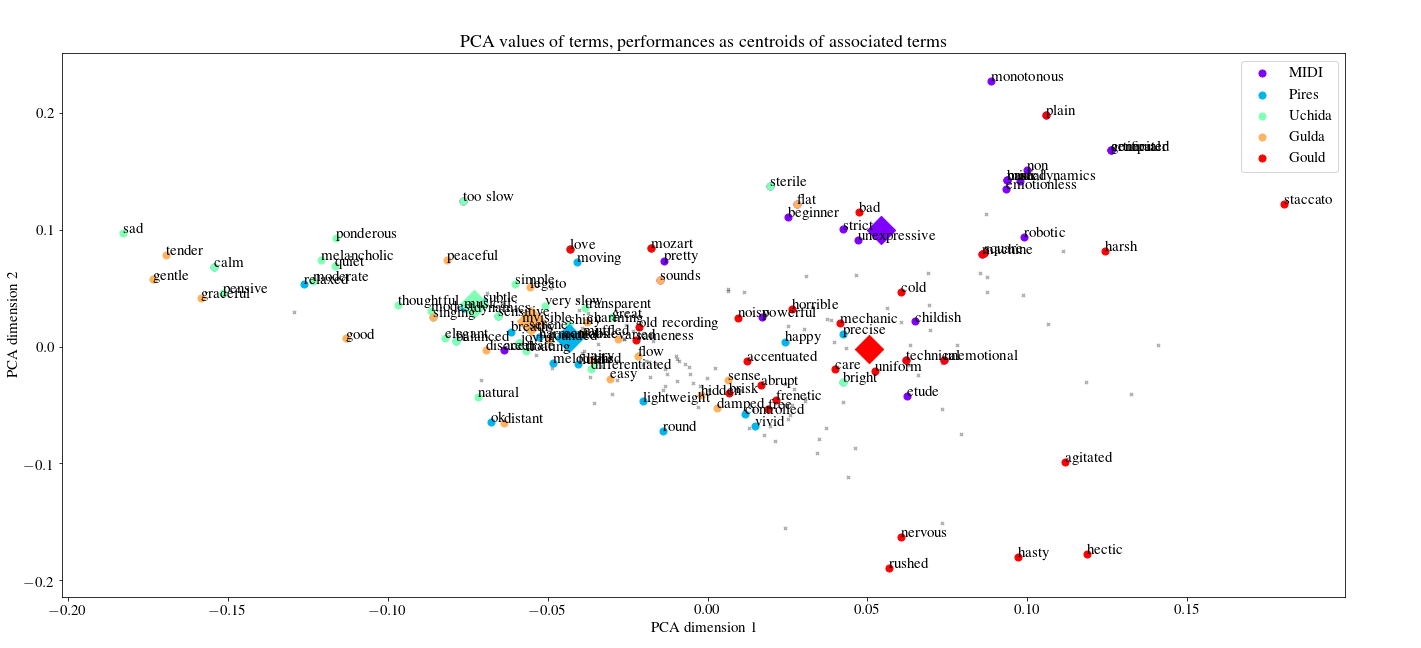}
\caption{A visualization of the first two dimensions recovered by the PCA. 
Dots represent terms used in the CEG. Colored terms from characterizations of Mozart's Sonata K 545, grey terms from other pieces. 
The terms are colored according to the performance they characterize.
}

\label{fig:PCA}
\end{figure*}

Table \ref{tab:expressive_dimensions} shows the terms that have the strongest loading on the dimensions in the above PCA. 
Dimension 1 carries intuitive meaning:
its extremes reach from \emph{`hectic'} and \emph{`agitated'} to \emph{`gentle'} and \emph{`calm'}.
The other three dimensions are harder to connect to a clear semantic dimension.
Note for instance the terms \emph{`cold'} and \emph{`warm'}; both influence dimension 4 strongly in the same direction.
Figure \ref{fig:PCA} displays the terms used to describe Mozart's Sonata in the space spanned by the dimensions 1 and 2.
The performances themselves are embedded in the space as the centroid of their respective terms.
Three clusters emerge, with the deadpan MIDI and Glenn Gould clearly sticking out,
and Mitsuko Uchida slightly more towards the \emph{`calm'} and \emph{`sad'} end of Dim.1 (an impression confirmed by listening).

\section{How do measurable/computable performance features relate to the expressive character dimensions?}\label{sec:q3_performance}
\begin{table*}[t]
\centering
\begin{small}
\begin{tabular}{@{}llllllll@{}}
\toprule
\multicolumn{2}{c}{Dimension 1}                         & \multicolumn{2}{c}{Dimension 2}                        & \multicolumn{2}{c}{Dimension 3}                        & \multicolumn{2}{c}{Dimension 4}                        \\
 \midrule
\multicolumn{2}{c}{PP ($R^2=0.24$)} & \multicolumn{2}{c}{PP ($R^2=0.18$)} & \multicolumn{2}{c}{PP ($R^2=0.26$)} & \multicolumn{2}{c}{PP	($R^2=0.24$)} \\
loudness avg                   & $0.51^{***}$           & loudness sk                & $0.45^{**}$               & loudness std               & $-0.53^{**}$              & beat period k                 & $-0.34^{*}$            \\
                               &                        &                            &                           &                            &                           & loudness std                  & $-0.44^{*}$            \\
\midrule
\multicolumn{2}{c}{MF ($R^2=0.39$)}     & \multicolumn{2}{c}{MF ($R^2=0.00$)}    & \multicolumn{2}{c}{MF ($R^2=0.00$)}    & \multicolumn{2}{c}{MF ($R^2=0.29$)}    \\
rhythmic complexity            & $-0.74^{*}$            & minorness                       & $0.15$                    & articulation               & $-0.15$                   & rhythmic complexity           & $0.52^{*}$             \\
tonal stability                      & $-0.94^{**}$           &                            &                           &                            &                           & tonal stability                     & $0.84^{***}$           \\
articulation                   & \ \ \ $0.46^{*}$             &                            &                           &                            &                           &                               &                        \\
\midrule
\multicolumn{2}{c}{HF $(R^2 = 0.22)$}  & \multicolumn{2}{c}{HF $(R^2 =0.00)$}  & \multicolumn{2}{c}{HF ($R^2 = 0.36$)} & \multicolumn{2}{c}{HF ($R^2 = 0.09$)} \\
valence sk                     & $0.48^{**}$            & valence avg                & $0.14$                    & valence k                  & $\ \ \ 0.42^{**}$               & valence k                     & $-0.33^{*}$            \\
                               &                        &                            &                           & arousal avg                & $-1.24^{***}$             &                               &                        \\
                               &                        &                            &                           & valence std                & \ \ \ $0.27^{*}$                &                               &                        \\
                               &                        &                            &                           & valence avg                & $-0.82^{*}$               &                               &                        \\ \bottomrule
\end{tabular}
\end{small}
\caption{Multiple Linear Regression Analysis.
PP, MF and HF refer to performance parameters, mid- and high- level features, respectively.
avg, std, k and sk denote average, standard deviation, kurtosis and skewness.
The values are the regression coefficients (indicating the contribution of that feature to the model).
$R^2$ is the adjusted coefficient of determination for the whole model.
$^*$, $^{**}$, and $^{***}$ indicate statistical significance at levels $\alpha=.05,.01$ and $\alpha<.001$, respectively. 
}
\label{tab:linear_regression}
\end{table*}

In this section we study whether there is a systematic relationship between the expressive character dimensions described in Section \ref{sec:q2_dimensions} and measurable or computed performance qualities that can be extracted directly from the audio files or from the score-to-performance alignments.
In the rest of this article, we refer to these measurable or computed performance qualities as \emph{performance features}.



\subsection{Description of Performance Features}\label{sec:audio_performance_features}
\subsubsection{Performance Parameters}

We consider two performance parameters, tempo and dynamics curves, to relate to the expressive dimensions described above.
The tempo curves are extracted directly from the hand-corrected score-to-performance alignments by computing the inter-beat intervals.
For computing loudness we use the loudness curve computed from the MIR Toolbox~\cite{MIRToolbox} using a perceptually weighted smoothing of the signal energy.
For inter- and intra-piece comparisons, we calculate the average value, standard deviation, kurtosis and skewness of these curves.
Average tempo/dynamics provides an indicator of how fast/loud a performance is, the standard deviation quantifies the tempo/loudness deviations, 
kurtosis provides a measure of how extreme these deviations are, and
skewness indicates how asymmetric the tempo/loudness values are (e.g., whether the piece is regularly faster or slower than the average tempo).

 \subsubsection{Mid-level Features}\label{sec:mid_level_features}
Mid-level features are perceptual qualities of music such as articulation, rhythmic clarity and modality that describe overall properties of musical excerpts and are intuitively clear to listeners~\cite{friberg2014using}. 
 We extract the seven mid-level features described in~\cite{aljanaki2018data}, using the deep convolutional network architecture from~\cite{chowdhury2019towards} (the A2Mid variant, specifically).
 The 7 mid-level features are \emph{melodiousness}, \emph{articulation}, \emph{rhythmic complexity}, \emph{rhythmic stability}, \emph{dissonance}, \emph{tonal stability}, and \emph{minorness} (see \cite{aljanaki2018data} for a detailed description of the features).
 We train our model on the mid-level features dataset~\cite{aljanaki2018data}, which contains 5000 audio snippets of 15 seconds each, and use the trained model to predict the mid-level features of the piano performances without any fine-tuning, as there is too little data for a supervised fine-tuning step. 
 To improve the validity of the transfer, we incorporate unsupervised domain adaptation~\cite{ganin2014unsupervised} during the training phase. 
 Since the pieces in the CEG are piano performances, we use a separate private collection of non-annotated piano music as the data source for domain adaptation. 
 We observe more variation in the mid-level predictions between the performances while using a domain-adapted model than a non-domain-adapted one, which indicates that it is a useful step in the pipeline.

 \subsubsection{High-level Features}\label{sec:high_level_features}
As high-level emotion-related descriptors, we choose the common \emph{arousal} and \emph{valence} dimensions~\cite{frijda1986emotions,Russell:1980wy} and aim to
predict these from the audio recordings,
to then relate them to the expressive character dimensions.
We train a dynamic arousal-valence prediction network using the DEAM dataset~\cite{aljanaki2017developing}. 
We tested a VGG-like model, similar to the one described in Section \ref{sec:mid_level_features}, and we observed that when the network is pre-trained on the mid-level dataset and extended with a multi-layer GRU-RNN (Gated Recurrent Unit Recurrent Neural Network) that is trained on the DEAM dataset, we get the best results. 
To improve the results further, we use the recently released receptive-field regularized ResNet~\cite{koutini2019receptive} for the pre-training phase, since it appears to give better results for short audio snippets than the VGG-like variant. 
The inputs to our network are Mel-spectrograms and we process 2-second segments of the spectrogram with 0.5-second hops.
As in the case of the expressive performances, in order to compare the predicted arousal and valence curves for inter-  and intra-piece comparisons, we compute
average, standard deviation, kurtosis and skewness of these curves for each performance of each piece.


\subsection{Analysis with Multiple Linear Regression}\label{sec:multiple_linear_analysis}
To study the relation between the performance parameters and mid- and high- level features to the expressive character dimensions described in Section \ref{sec:q2_dimensions} we use multiple linear regression (MLR) analysis.
In this analysis, the \emph{dependent variables} are each of the expressive character dimensions (Dimensions 1 to 4) and the \emph{independent variables} are the performance features described above.
We carry out $4 \times 3 = 12$ MLRs for each expressive character dimension (4 in total) and subset of performance features (expressive parameters, mid- and high-level features).
Each of these regressions investigates whether each subset of performance features (expressive parameters, mid-/high-level) can significantly predict the position of the pieces in the expressive character dimensions.
The position of each piece in the 4D expressive character space is computed as the centroid of all of its terms in this space.
For each of these MLRs we perform a variable selection using the Zheng-Loh method~\cite{ZhengLoh:1997}.
The results are summarized in Table \ref{tab:linear_regression}.
The MLR results indicate that the expressive parameters are significant predictors of all 4 expressive character dimensions, 
with medium effect sizes ($R^2$).
Mid-level features are only significant predictors of Dimensions 1 and 4. 
High-level features are only significant for Dimensions 1 and 3. 
Thus, Dimension 1 (the \emph{`gentle'}/\emph{`calm'} vs.~\emph{`hectic'}/\emph{`agitated'} axis, see Section~\ref{sec:q2_dimensions}) seems systematically related to our performance features
at all three levels, which further corroborates its significance.

\section{Discussion}\label{sec:discussion} 

In Section \ref{sec:readability} we observed a small positive relationship between the complexity of verbal descriptions and listeners' musical training. 
We expect that stronger evidence of a relationship would emerge if musical training were better controlled for (our sample had few listeners with $<5$ years of training) and the complexity measure were further developed to account for specialized musical terms. 
Our analysis of listeners' preferred performances in Section \ref{sec:listeners_preference} revealed that the deadpan performances and performances by Glenn Gould were least well-liked.
Prior research has suggested that listeners prefer quantitatively average expressive performances~\cite{Repp1997}, which might explain partially the lack of enthusiasm for Gould's idiosyncratic playing. 

The results in Sections \ref{sec:semantic_similarity} and \ref{sec:q2_dimensions} suggest that listeners tend to describe performances of the same piece similarly, although there is  some variability (e.g., a performance can be described both as \emph{`beautiful'} or \emph{`bad'} by different listeners; cf. both \emph{`cold'} and \emph{`warm'} being negatively correlated with Dimension 4).
An important issue is that NLP methods for assessing similarity between the descriptions are not really suitable for analyzing performance descriptions, where each term is loaded with complex meaning\footnote{For example, the performance of the Mozart piece by Austrian-trained Japanese pianist Mitsuko Uchida was described by a participant as `Russian pianist'. To understand this description, it is necessary to have the concept of the Russian School of performance.} as well as many cross-domain mappings (e.g., metaphors).

The results in Section \ref{sec:multiple_linear_analysis} reveal relationships between performance features and expressive dimensions that conform to musical intuition, with the effects being most pronounced for expressive character Dimension 1 (which is also the one that we find easiest to interpret, see table \ref{tab:expressive_dimensions}).
For instance, the analysis suggests that louder performances or performances with large outliers in the valence curve would be perceived as more irregular and agitated, while softer performances or performances without large outliers in valence would be perceived as calm or graceful.


\section{Conclusions and Future Work}\label{sec:conclusions}
This paper has introduced the CEG dataset
and presented some exploratory analysis addressing three main questions related to
inter-listener agreement, main emerging description dimensions, and relations between
user characterizations and measurable performance parameters.


Future work will focus on a more in-depth analysis of the question of semantic similarity.
As discussed in Section \ref{sec:q2_dimensions}, the description of expressive character includes many nuances that are not well suited to be analyzed with generic NLP methods, given how loaded with meaning certain terms are.
We plan to investigate methods like \emph{pile sorting}~\cite{Trotter:Pilesorting}
with expert musicians to devise a meaningful semantic clustering of the terms.
Furthermore, we plan to collect more human annotations (e.g., mid- and high-level features)
as a basis for a more systematic comparison.

\section{Acknowledgments}

This research has received support from the European Research Council (ERC) under the European Union’s Horizon 2020 research and innovation programme under grant
agreement No. 670035 (project ``Con Espressione'') and by the Research Council of Norway through its Centers of Excellence scheme, project number 262762 and the MIRAGE project, grant number 287152.
We gratefully acknowledge the effort invested by our music expert,
Hans Georg Nicklaus (Anton Bruckner Private University of Music, Linz) for helping with the selection of the different performances in the dataset.
We thank Olivier Lartillot for sharing the Matlab code to compute the loudness features.

%


\bibliography{bibliography}

%
%
%
%

\end{document}